\pdfoutput=1
\documentclass[
rsi,
amsmath,amssymb,
reprint,
showkeys,
showpacs
]{revtex4-1}
\usepackage{amssymb}
\usepackage{amsthm}
\usepackage{amsmath}
\usepackage{graphicx}
\usepackage{dcolumn}
\usepackage{bm}
\usepackage{color}
\usepackage{float}
\usepackage{natbib}
\usepackage{subcaption}
\usepackage{textcomp}
\usepackage{hyperref}

\begin{document}
\title{Ultracold Neutron Properties of the Eljen-299-02D deuterated scintillator}
\author{Z.~Tang}
\email{ztang@lanl.gov}
\affiliation{Los Alamos National Laboratory, Los Alamos, New Mexico 87545, USA}
\author{E.~B.~Watkins}
\affiliation{Los Alamos National Laboratory, Los Alamos, New Mexico 87545, USA}
\author{C.~Cude-Woods}
\altaffiliation{Los Alamos National Laboratory, Los Alamos, New Mexico 87545, USA}
\affiliation{North Carolina State University,  Raleigh, North Carolina 27695, USA}
\author{S.~M.~Clayton}
\affiliation{Los Alamos National Laboratory, Los Alamos, New Mexico 87545, USA}
\author{J.~H.~Choi}
\affiliation{North Carolina State University,  Raleigh, North Carolina 27695, USA}
\author{S.~A.~Currie}
\affiliation{Los Alamos National Laboratory, Los Alamos, New Mexico 87545, USA}
\author{F.~Gonzalez}
\affiliation{Indiana University, Bloomington, Indiana 47405, USA}
\author{D.~E.~Fellers}
\affiliation{Los Alamos National Laboratory, Los Alamos, New Mexico 87545, USA}
\author{N.~C.~Floyd}
\altaffiliation{Los Alamos National Laboratory, Los Alamos, New Mexico 87545, USA}
\affiliation{University of Kentucky, Lexington, Kentucky 40506, USA}
\author{Md.~T.~Hassan}
\affiliation{Los Alamos National Laboratory, Los Alamos, New Mexico 87545, USA}
\author{K.~P.~Hickerson}
\affiliation{W. K. Kellogg Radiation Laboratory, California Institute of Technology, Pasadena, California 91125, USA}
\author{A.~T.~Holley}
\affiliation{Tennessee Technological University, Cookeville, Tennessee 38505, USA}
\author{D.~E.~Hooks}
\affiliation{Los Alamos National Laboratory, Los Alamos, New Mexico 87545, USA}
\author{T.~M.~Ito}
\affiliation{Los Alamos National Laboratory, Los Alamos, New Mexico 87545, USA}
\author{B.~A.~Johnson}
\altaffiliation{Los Alamos National Laboratory, Los Alamos, New Mexico 87545, USA}
\affiliation{Utah State University, Logan, Utah 84322, USA}	
\author{J.~C.~Lambert}
\altaffiliation{Los Alamos National Laboratory, Los Alamos, New Mexico 87545, USA}
\affiliation{Utah State University, Logan, Utah 84322, USA}	
\author{S.~K.~Lawrence}
\affiliation{Los Alamos National Laboratory, Los Alamos, New Mexico 87545, USA}
\author{C.~Y.~Liu}
\affiliation{Indiana University, Bloomington, Indiana 47405, USA}
\author{S.~W.~T.~MacDonald}
\affiliation{Los Alamos National Laboratory, Los Alamos, New Mexico 87545, USA}
\author{M.~Makela}
\affiliation{Los Alamos National Laboratory, Los Alamos, New Mexico 87545, USA}
\author{C.~L.~Morris}
\affiliation{Los Alamos National Laboratory, Los Alamos, New Mexico 87545, USA}
\author{L.~P.~Neukirch}
\affiliation{Los Alamos National Laboratory, Los Alamos, New Mexico 87545, USA}
\author{A.~Saunders}
\affiliation{Los Alamos National Laboratory, Los Alamos, New Mexico 87545, USA}
\author{C.~M.~O'Shaughnessy}
\affiliation{Los Alamos National Laboratory, Los Alamos, New Mexico 87545, USA}
\author{R.~W.~Pattie}
\affiliation{East Tennessee State University, Johnson City, Tennessee 37614, USA}
\author{A.~R.~Young}
\affiliation{North Carolina State University,  Raleigh, North Carolina 27695, USA}
\author{B.~A.~Zeck}
\altaffiliation{Los Alamos National Laboratory, Los Alamos, New Mexico 87545, USA}
\affiliation{North Carolina State University,  Raleigh, North Carolina 27695, USA}

\begin{abstract}
In this paper we report studies of the Fermi potential and loss per bounce of ultracold neutron (UCN) on a deuterated scintillator (Eljen-299-02D). These UCN properties of the scintillator enables a wide variety of applications in fundamental neutron research.    
\end{abstract}


\maketitle

\section{Introduction}
\begin{figure}
\includegraphics[scale=0.54]{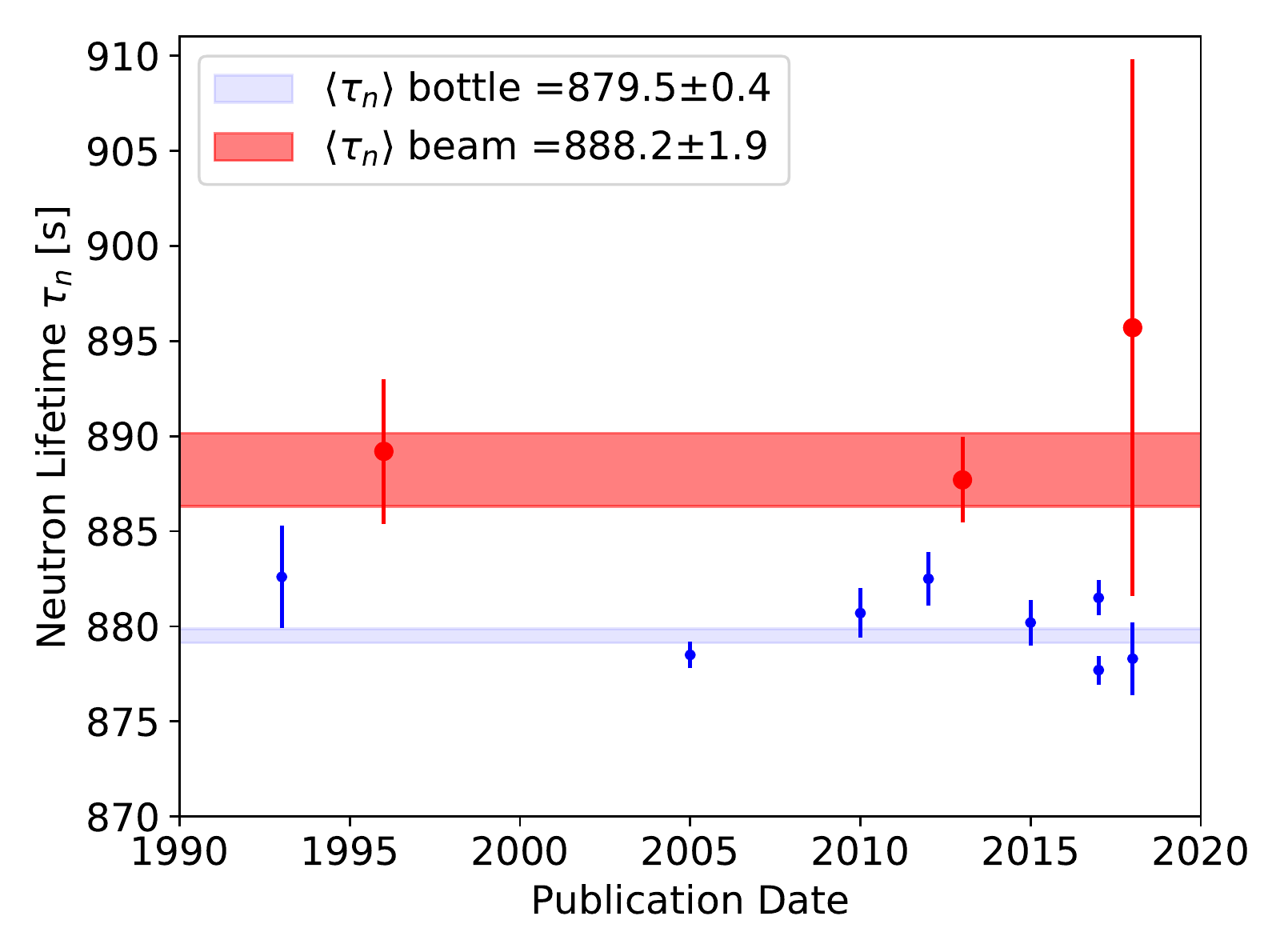} 
\caption{Red/blue points are the data with error bars for recent beam/bottle neutron lifetime experiments. The red/blue bands are the average values for the two methods with $\pm\sigma$ error bands}
\label{fig:lifetime}
\end{figure}

In the Standard Model of particle physics, the free neutron decay ($n \rightarrow p+e+\bar{\nu}_e$) has a characteristic lifetime $\tau_n$ of about 15 minutes. There are two different methods for measuring $\tau_n$: experiments that measure the decay rate of neutrons in cold neutron beams \cite{Byrne1996, Nico2005, Yue2013} or experiments that measure the survival of bottled ultracold neutrons (UCN) \cite{Serebrov2008, Pattie2018, Ezhov2018}. By counting the number of protons emitted from neutron beta decay in a well-calibrated cold neutron beam, the beam method measures the mean time for neutrons to decay into protons, with an averaged result of 888.1 $\pm$ 2.0 s. In the Standard Model, this time is equivalent to the total neutron lifetime with the exception of the rare process of neutrons decaying into bound hydrogen atoms and electron antineutrinos, which has a calculated branching ratio of $4 \times 10^{-6}$ \cite{Kabir1967, Nemenov1980,Song1987}. The bottle experiments utilize trapped UCN, which are neutrons with kinetic energy less than 350 neV. At this energy, the UCN can undergo total external reflection on material walls, and their kinetic energy is on the same scale as their gravitational and magnetic potential energies. The UCN bottle experiments utilize these UCN properties to trap the neutrons and measure numbers that remain after a certain storage time, with an average lifetime result of 879.5 $\pm$ 0.4 s. These two methods differ by 8.7 seconds, or 4.5 standard deviations (Fig.~\ref{fig:lifetime}). 

Recently, some authors have suggested the possibility of 
hidden decay/oscillation channels of the neutron decay that have so far eluded detection \cite{Berezhiani2009,Fornal2018}. Although many experiments have eliminated some of these decay channels \cite{Tang2018a,Sun2018,McKeen2018,Berezhiani2018}, some parameter space still remains \cite{McKeen2018,Cline2018}. We propose a new experiment \cite{Tang2018b} to measure the neutron beta decay lifetime by measuring number of neutrons and number of beta decays similar to the beam lifetime experiments. 
The main difference in the proposed experiment is that we will measure the electrons from beta decay as opposed to the protons, which will be subjected to entirely different systematic effects. The experiment will use a bottle made of deuterated polystyrene (d-PS) based scintillator to trap the UCN and simultaneously measure the electrons from beta decay. In this paper, we report a study of the Fermi potential and loss per bounce properties of an Eljen-299-02D deuterated scintillator. Such a scintillator could also be used in other UCN related experiments due to its UCN storage properties.

\section{Fermi Potential}
\begin{figure}
\includegraphics[angle=-90, scale=0.33]{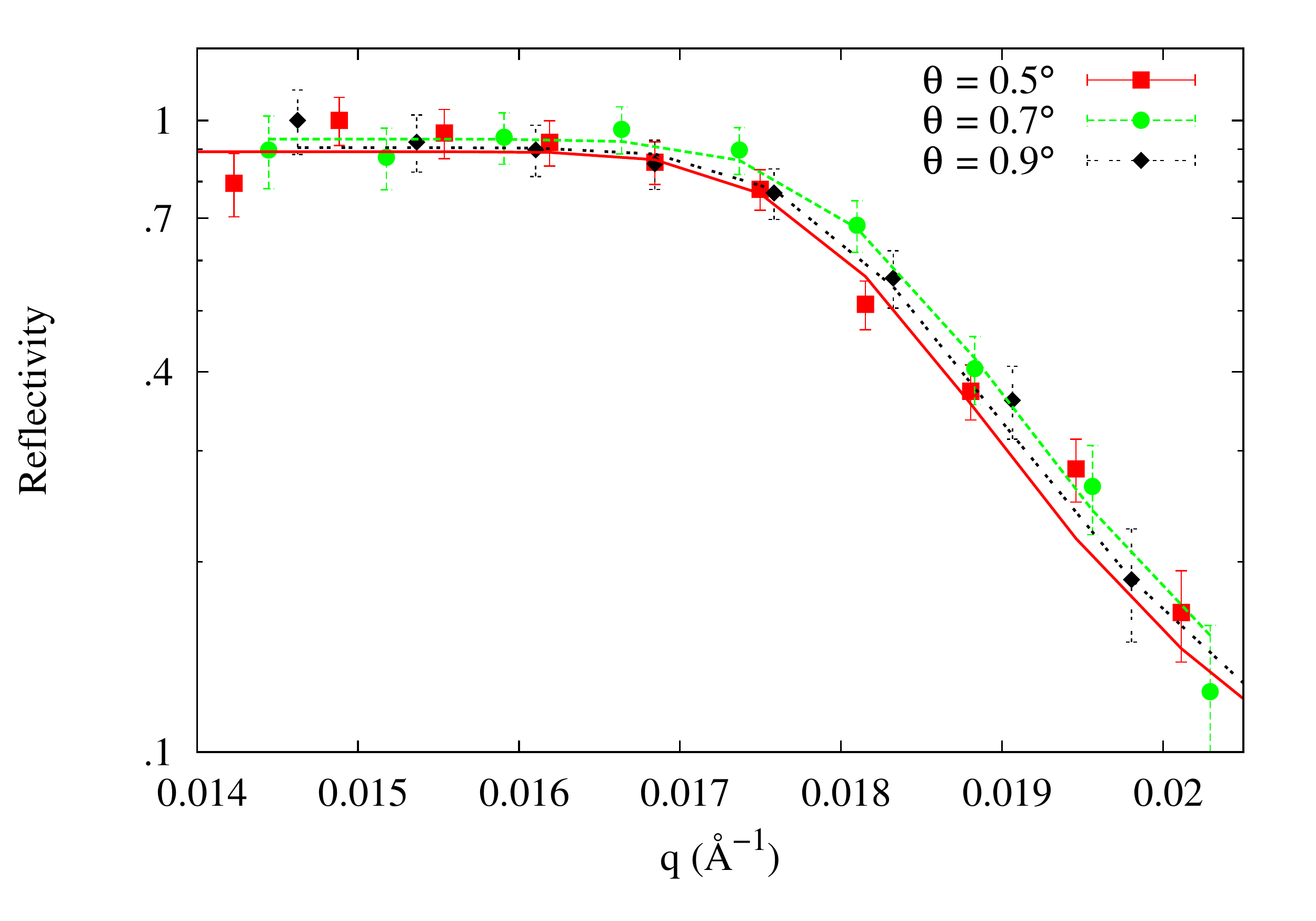}
\caption{Neutron reflectometry data (symbols) and fits (lines) to d-PS for three sets of incident angles. An average scattering length density, $\beta$, was obtained by fitting the Fresnel reflectivity function to the multiple data sets. Three measurements using different incident angles of the neutrons were used to ensure the accuracy of the fitted $\beta$ value. }
\label{Fig:refl}
\end{figure}

The Fermi potential of d-PS was measured using Asterix, a time of flight neutron reflectometer at the Los Alamos Neutron Scattering Center (LANSCE)\cite{Watkins2016}. 
Asterix views a liquid H$_2$ moderator providing a polychromatic cold neutron beam with wavelengths, $\lambda$, ranging from 4 to 13 \AA. 
The neutron beam divergence and spot size on the sample were controlled by two sets of collimating slits. 
Reflectivity, R($q$), is defined as the ratio of the intensity of the reflected beam to the incident beam as a function of the neutron momentum transfer vector normal to the reflecting surface, $q$, where $q=4\pi \sin(\theta)/\lambda$. 
Total external reflection was measured up to a critical momentum transfer, $q_c= \sqrt{16\pi\beta}$ , where $\beta$ is the scattering length density of the sample. The relationship between $\beta$ and the Fermi potential, $V_F$, is given in Eqn.~\ref{Eqn:fermi}, where $m_n$ is the mass of the neutron. 
\begin{equation}
V_F =\dfrac{2\pi\hbar^2}{m_n}\beta
\label{Eqn:fermi}
\end{equation}
Multiple neutron reflectometry experiments were performed to measure $q_c$ using the polychromatic beam with incidence angles on the sample, $\theta$, ranging from approximately 0.5\textdegree to 0.9\textdegree and approximately 10\% $dq/q$ resolution. Scattered neutrons were collected using a 2D $^3$He position sensitive detector as a function of $\theta$ and $\lambda$ which simultaneously captured both the specular reflectivity signal and off-specular scattering originating from the surface roughness. High surface roughness of the sample yielded intense off-specular scattering, and subtraction of the off-specular scattering signal from the specular reflection, R(q),  was a significant contribution to the uncertainty in fitting the data. To mitigate the influence of background subtraction on the value of the Fermi potential, multiple measurements were made using different incident angles of the neutron beam on the sample (Fig~\ref{Fig:refl}). The data was fit using Fresnel’s law for reflection from an ideal interface which captures the total external reflection of neutrons up to $q_c$ followed by a $q^{-4}$ drop in intensity. A normalization factor and the $\beta$ for d-PS were the only free parameters used in the fit, and error estimates on the $\beta$ parameter were based on a $\chi^2 + 1$ metric. Values obtained for $\beta$ from fitting three independent measurements were $6.42 \pm 0.08 \times 10^{-6} \rm{\AA^{-2}}$, $6.53 \pm 0.12 \times 10^{-6} \rm{\AA^{-2}}$, and $6.48 \pm 0.11 \times 10^{-6} \rm{\AA^{-2}}$ corresponding to Fermi potentials of $167.2 \pm 2.1$ neV, $170.0 \pm 3.1$ neV, and $168.7 \pm 2.9$ neV. Averaging the three measurements yields a $\beta$ of $6.48 \pm 0.06 \times 10^{-6} \rm{\AA^{-2}}$ corresponding to a Fermi potential of $168.2 \pm 1.5$ neV. 

\section{Scintillator loss-per-bounce measurement}
\begin{figure}[!h]
\centering
\includegraphics[scale=1.35]{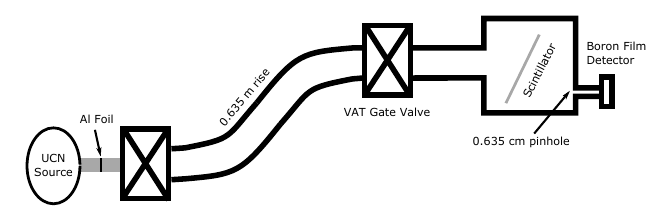}
\caption{Schematic diagram for the loss-per-bounce measurement}
\label{Fig:LPB}
\end{figure}

\begin{figure}[!h]
\centering
\hspace*{-0.63cm}
\includegraphics[scale=0.5]{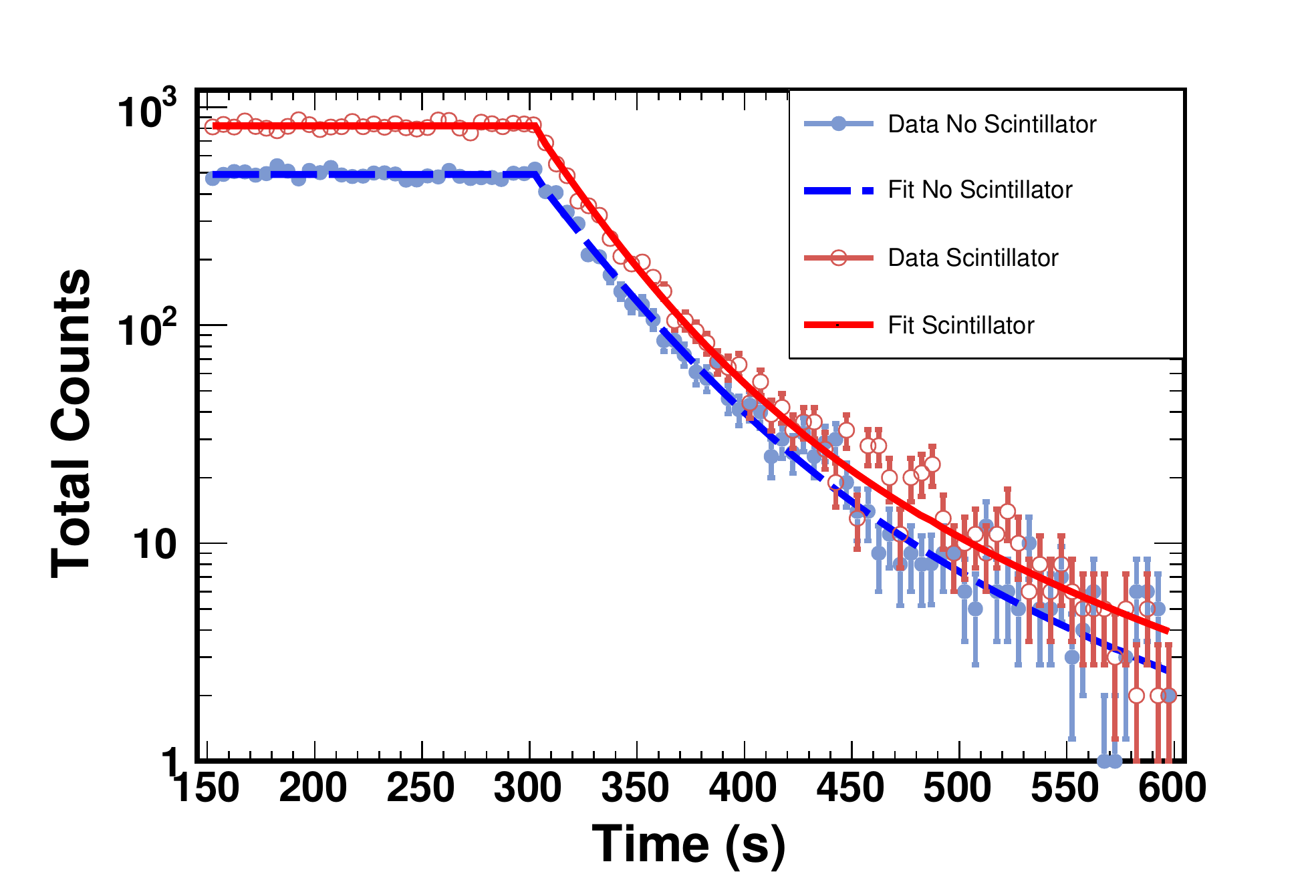} 
\caption{Histogram of the total number counts seen by the pinhole detector for the two different run configurations as a function of time. The solid and dashed lines represent the measurements with and without the deuterated scintillator respectively. A global fit of the two data sets was performed using two free parameters, loss per bounce of the stainless-steel bottle, and loss factor, $f$, of the scintillator. The equilibrium plateau of each configuration (150 s to 300 s) was used to fit for the normalization factor of each data set. }  
\label{Fig:fits}
\end{figure}

A loss-per-bounce measurement for the Eljen-299-02D scintillator was performed by measuring the lifetimes of a stainless steel UCN bottle with and without the scintillator placed inside the bottle. The volume and surface area of the bottle are 3270 cm$^3$ and 1350 cm$^2$ respectively, and the surface area of the scintillator is 292 cm$^2$. The UCN bottle was connected to a port off the Los Alamos UCN source \cite{Ito2018} and was separated from the deuterium volume by a 0.001" thick 1100 series aluminum foil. The bottle was raised 0.635 meters to account for the change in Fermi potential between the Nickel-Phosphorus coated source guide \cite{Pattie2017} and the scintillator. The UCN are loaded into the bottle for 300 seconds with the upstream gate valves open. Once the UCN density is well saturated inside the bottle, the gate valves are then closed, and the lifetime curve of the bottle is extracted by monitoring the rate of UCN loss through a 0.635 cm diameter pinhole boron film detector \cite{wang2015} (Fig.~\ref{Fig:LPB}).  The lifetime curves with and without the scintillator are shown in Fig.~\ref{Fig:fits}.

The analysis for the loss per bounce of the scintillator was performed simultaneously using the two data sets: the loss per bounce of the stainless steel UCN bottle is a fitting parameter for both data sets, and the loss per bounce of the scintillator is only relevant for the data set with the scintillator. The spectral evolution model used to fit the two data sets is described in Eqn.~\ref{Eqn:LPB}, where $N(t)$ is the number of UCN in the trap as a function of a time, $\rho(E)$ is the initial energy spectrum of the UCN, and $\tau(E)$ is the bottle lifetime as a function of energy. An additional velocity weight is added to account for the preferential detection of faster UCN due to the pinhole detector. $\tau_\beta^{-1}$, $\tau_{ss}^{-1}$, $\tau_{pinhole}^{-1}$, and $\tau_{scint}^{-1}$ represent the UCN loss rates due to neutron lifetime, stainless steel bottle, pinhole detector, and scintillator respectively. 
  
\begin{equation}
\begin{split}
N(t) &= \int^{E_{max}}_{E_{min}} \rho (E) v e^{-t/\tau(E)} dE \\
\tau(E)^{-1} &= \tau_\beta^{-1} + \tau_{ss}(E)^{-1} + \tau_{pinhole}(E)^{-1} + \tau_{scint}(E)^{-1}\\
\label{Eqn:LPB}
\end{split}
\end{equation}  

In this paper, we used the initial velocity distribution as outlined in  \citep{Pattie2017, Pattie2020}, where the collision rate weighted velocity distribution is  $\rho v \sim v^{2.7}$. This initial spectrum is then adjusted to account for the height difference and loading time as shown in Eqn.~\ref{Eqn:rise}, where $E$ and $E_{rise}$ are the kinetic energies of the UCN and the rise in the height of the beamline (64.8 neV) respectively. Here we have used an energy range of 64.8 neV to 186 neV for the initial spectrum, which is then adjusted by the relative height of the beamline. This initial spectrum is then sliced into 31 equal time bins and evolved to $t = 300$ seconds. The 31 time bins were chosen due to the of the repetition rate (0.1 Hz) of the UCN source.

\begin{equation}
\begin{split}
\rho(E,h=0,t=0)~dE &= E^{0.35}~dE \\
\rho(E,h=0.635 \rm{m},t=0)~\textit{dE} &= E^{0.35} \sqrt{\frac{E-E_{rise}}{E}}~dE \\
\rho(E,h=0.635 \rm{m}, t =300)~\textit{dE} &= \sum\limits_{i=0}^{30} \frac{1}{31}E^{0.35} \sqrt{\frac{E-E_{rise}}{E}} \\
&\times e^{-10i/\tau(E)}~dE
\end{split}
\label{Eqn:rise}
\end{equation} 

In our model, we assume that the initial velocity directions of the UCN are sufficiently mixed that the kinetic theory for non-interacting gas applies in the calculation of the wall interaction rate (Eqn.~\ref{Eqn:loss_mech}), where $A$ is the total surface area of the material, $|v|$ is the magnitude of the UCN velocity, and $U$ is the volume of the bottle. A loss-per-bounce parameter, $\mu$, is also added onto this equation to obtain the loss rate for each component.
\begin{equation}
\frac{1}{\tau_i} = \frac{A_i |v| \mu_i}{4 U} 
\label{Eqn:loss_mech}
\end{equation}
For $\mu_{ss}$, we assumed an energy independent loss-per-bounce parameter due to the combination of losses due to gaps in the system and the losses on the surface.  For $\mu_{pinhole}$, we assumed a loss rate of unity due to the UCN detector, which is a vaild assumption since the surface area for the detector is 64 times larger than the surface area of the pinhole.
For $\mu_{scint}$, an energy dependent loss-per-bounce model that integrates over all incident angles is used (Eqn.~\ref{Eqn:ffactor})\cite{golub1991ultra}, where $V_F$ is the Fermi potential of the scintillator, $E$ is the kinetic energy of the UCN, and $f$ is the ratio of the imaginary to the real part of the Fermi potential. The fit for the loss per bounce in stainless steel and $f$ for the scintillator is given in Fig.~\ref{Fig:fits}.

\begin{equation}
\mu_{scint}(E) = 2f\bigg[\dfrac{V_F}{E}\sin^{-1}\sqrt{\dfrac{E}{V_F}}-\sqrt{\dfrac{V_F}{E}-1}\bigg]
\label{Eqn:ffactor}
\end{equation}   

\begin{figure}
\centering
\hspace*{-0.4cm}
\includegraphics[scale=0.56]{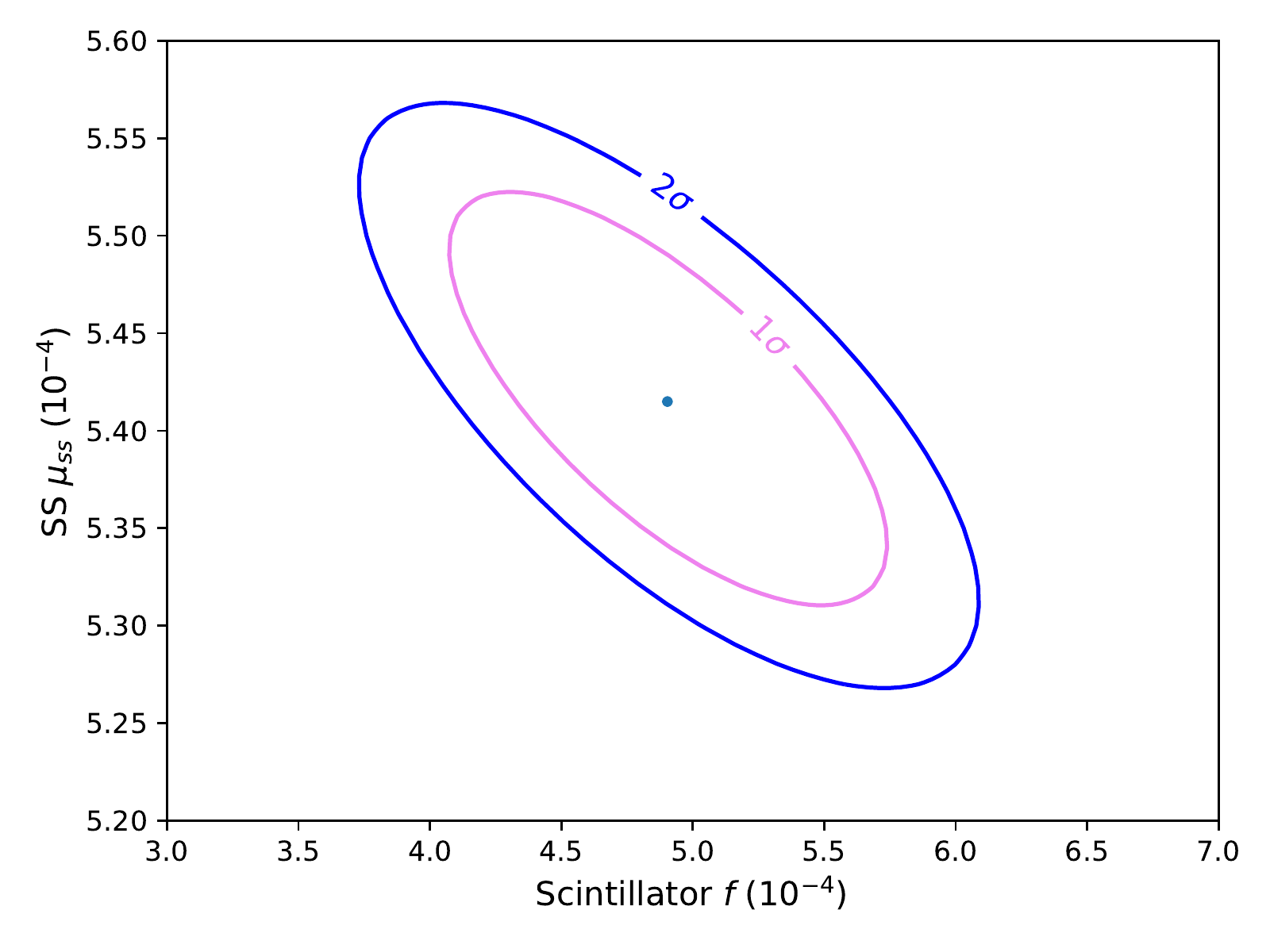}
\caption{Contour map showing the $\chi^2$ minimum of the global fit as a function of $f$ and $\mu_{ss}$. The reduced chi-squared, $\chi^2/\nu$, is 1.0. The error ($\sigma$) in the global fit was determined using the $\chi^2+1$ regions.}
\label{Fig:Contour}
\end{figure}

We have used an iterative approach in our analysis: first, initial guesses of $f$ and $\mu_{ss}$ are used to evolve the UCN energy spectrum to 300 seconds; then, that energy spectrum is used as an input into a global chi-squared minimization of two data set. The iteration is complete when the initial guess values for $f$ and $\mu_{ss}$ match the central values from the minimization, yielding a result of $f=4.9\pm0.8\times 10^{-4}$ for the scintillator and an energy independent loss per bounce for stainless steel of $\mu_{ss}=5.4\pm0.1 \times 10^{-4}$ with $\chi^2/\nu = 1.0$. The error in the global fit was determined by taking the limits of the $\chi^2+1$ region for the fit as shown in Fig.~\ref{Fig:Contour}. 

\section{Surface Roughness Measurements}
\begin{figure} [!h]
\begin{subfigure}{.5\textwidth}
\centering
\includegraphics[width=0.75\textwidth]{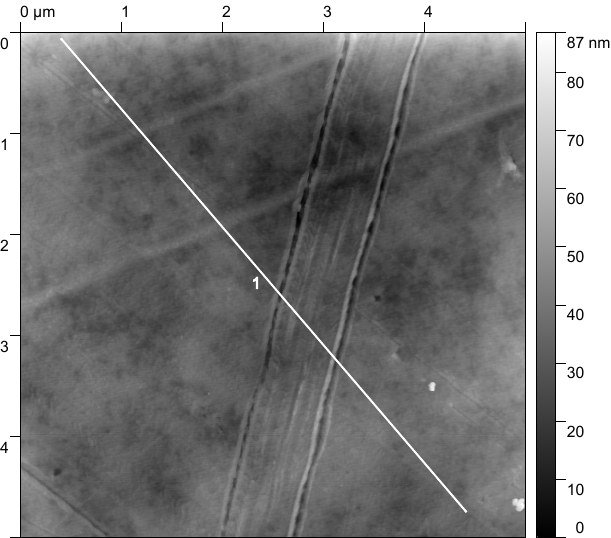}
\caption{}
\end{subfigure}
\begin{subfigure}{.5\textwidth}
\centering
\hspace{-1.5cm}
\includegraphics[width=0.71\textwidth]{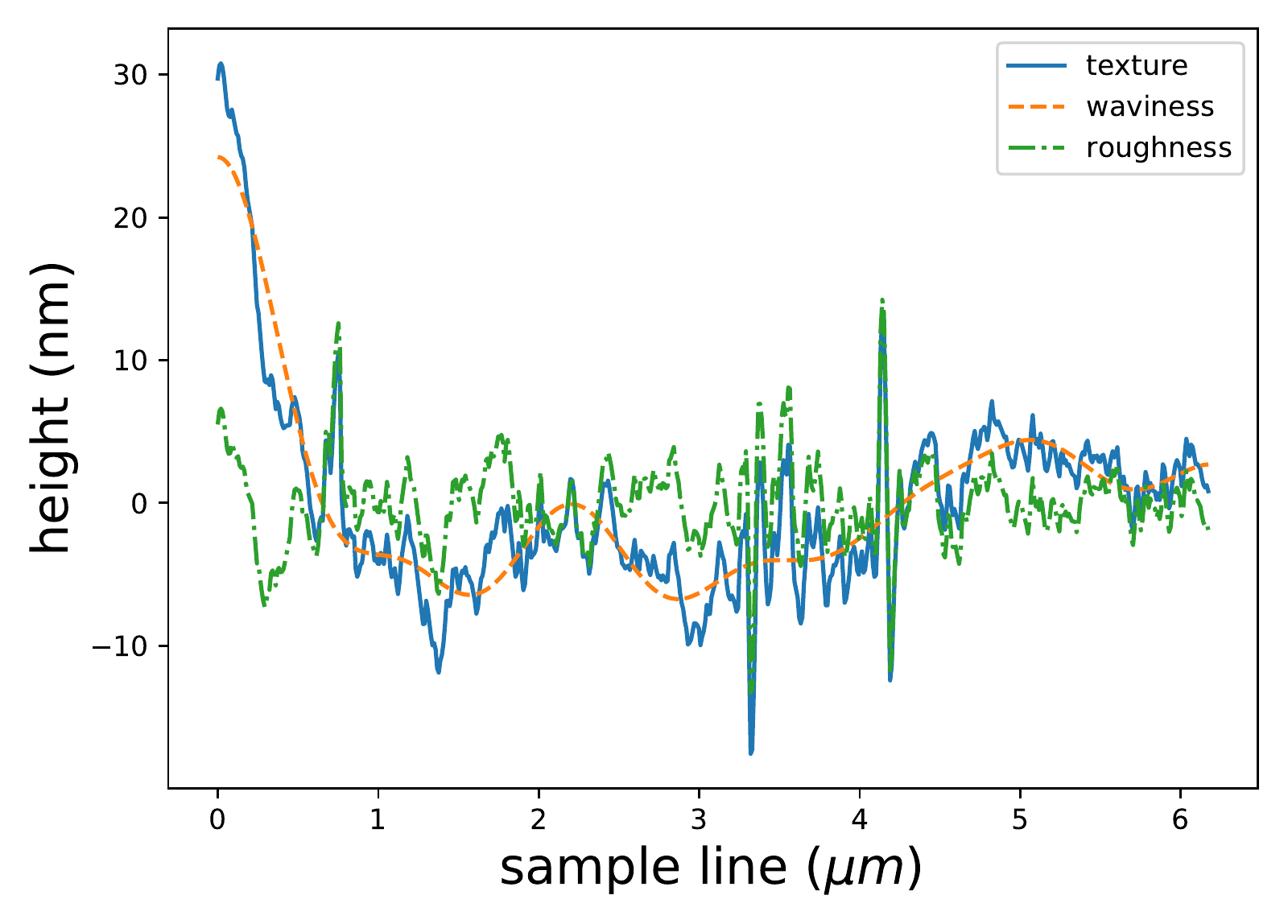}
\caption{}
\end{subfigure}
\caption{(a) Sample profilometry data of a $10 \mu m \time 10 \mu m$ of the deuterated polystyrene scintillator. (b) Lineout from left to right as indicated by the white line in (a). The waviness was subtracted from data (texture) to obtain the roughness plot. }
\label{fig:rough}
\end{figure}

The calculated value of the loss factor using the manufacturer-provided elemental and isotopic composition (97 \% deuterium purity) is $1\times10^{-4}$, which is smaller than the measured value of $4.9\pm0.8 \times 10^{-4}$. 
The hydrogen impurity in the scintillator is due to the isotopic purity of deuterated styrene monomers and the hydrogen present in the primary fluorescent emitter.
One possible explanation is that the surface is rough at or below the UCN wavelength scale. In this case, the loss-per-bounce parameter, $\mu$, will be modified due to the modification of the local Fermi potential \cite{golub1991ultra}. Eqn.~\ref{Eqn:mu_rough} shows the effect on the loss-per-bounce parameter, where $k_c = (2mV_F/\hbar^2)$ is the critical wavelength for the neutron, $\sigma$ is the RMS height variation of the surface roughness, and $w$ is the correlation length of the roughness.
\begin{equation}
\mu' = \mu \sqrt{1 + \dfrac{2\sigma^2k_c^2}{1+0.85k_c w+2k_c^2w^2}}
\label{Eqn:mu_rough}
\end{equation}
A measurement of the surface roughness of the scintillator was performed using a Bruker Dimension Icon atomic force microscope. The instrument was operated in peakforce tapping mode with standard ScanAsyst-Air tip. We sampled multiple 10 $\mu m$ x 10 $\mu m$ spots on the scintillator and obtained surface roughness by fitting to line outs (Fig.~\ref{fig:rough}). Since the loss-per-bounce parameter is only affected by roughness comparable to the wavelength of neutron, we estimated the surface roughness by separating out the long wavelength waviness features from the short wavelength roughness. The analysis was performed using Gwyddion, a scanning probe microscopy analysis software. We found the average roughness ranges from 2 - 8 nm, and the correlation length ranges from 40 - 200 nm, which is only a 2 \% correction on the loss-per-bounce parameter in the worst case. These results indicate that the surface roughness of the scintillator is not enough to account for the differences.

\section{Conclusion}
The Fermi potential for the Eljen-299-02D scintillator was measured using a neutron reflectometry beamline at LANSCE (Asterix) with a value of $168.2 \pm 1.5$ neV, consistent with the calculated value of 165.8 neV.  
The measured loss factor of $4.9\pm0.8 \times 10^{-4}$ did not agree with the calculated value of $1\times10^{-4}$, and this result cannot be explained by the roughness of the scintillator surface. Similar anomalous UCN losses have been explain by \onlinecite{Korobkina2004}, where water absorption on the surface of the metallic surfaces was the culprit. 

Regardless of the origin of the anomalous loss, the loss factor measurement did demonstrate the utility of the scintillator in trapping UCN and detecting electrons from neutron beta decay simultaneously, which can be used in future UCN based ``beam" lifetime and beta decay correlation experiments.
The measured loss factor is also sufficient for the statistical sensitivity that we would like to achieve for the proposed neutron beta decay lifetime experiment using a d-PS scintillator bottle. Taking the UCN spectrum and density obtained from Ref.~\citep{Ito2018}, we estimate the decays per measurement cycle to be between 2180 to 5640 events for a nominal 2 liter volume, with an observation time of 134 s and 206 s respectively. The storage times for the upper- and lower-bound estimates were determined using the time when decay counts is equal to the remaining UCN population. To obtain an one second statistical sensitivity on the neutron beta decay lifetime will require approximately 180 to 460 measurement cycles. 

\section*{Acknowledgments}
This work was supported by the Los Alamos National Laboratory LDRD Program (Project No. 20190048ER), National Science Foundation (grant no. PHY1914133), and U.S. Department of Energy, Office of Nuclear Physics (grant no. DE-FG02-97ER41042). 
This work was performed, in part, at the Center for Integrated Nanotechnologies, an Office of Science User Facility operated for the U.S. Department of Energy (DOE) Office of Science.
We gratefully acknowledge the support provided by the LANL Physics, AOT, MPA, and Sigma divisions. We would like to thank Eljen Technology and Charles Hurlbut for their technical support and useful discussions.

\bibliography{library}
\end{document}